# QUANTITATIVE ANALYSIS OF THE POTENTIAL ROLE OF BASAL CELL HYPERPLASIA IN THE RELATIONSHIP BETWEEN CLONAL EXPANSION AND RADON CONCENTRATION

Emese J. Drozsdik, Balázs G. Madas[*]
Radiation Biophysics Group, Environmental Physics Department, MTA Centre for Energy Research, Konkoly-Thege Miklós út 29-33, Budapest 1121, Hungary

Applying the two-stage clonal expansion model to epidemiology of lung cancer among uranium miners, it has been revealed that radon acts as a promoting agent facilitating the clonal expansion of already mutated cells. Clonal expansion rate increases non-linearly by radon concentration showing a plateau above a given exposure rate. The underlying mechanisms remain unclear. Earlier we proposed that progenitor cell hyperplasia may be induced upon chronic radon exposure. The objective of the present study is to test whether the induction of hyperplasia may provide a quantitative explanation for the plateau in clonal expansion rate. For this purpose, numerical epithelium models were prepared with different number of basal cells. Cell nucleus hits were computed by an own-developed Monte-Carlo code. Surviving fractions were estimated based on the number of cell nucleus hits. Cell division rate was computed supposing equilibrium between cell death and cell division. It was also supposed that clonal expansion rate is proportional to cell division rate, and therefore the relative increase in cell division rate and clonal expansion rate are the same functions of exposure rate. While the simulation results highly depend on model parameters with high uncertainty, a parameter set has been found resulting in a cell division rate exposure rate relationship corresponding to the plateau in clonal expansion rate. Due to the high uncertainty of the applied parameters, however, further studies are required to decide whether the induction of hyperplasia is responsible for the non-linear increase in clonal expansion rate or not. Nevertheless the present study exemplifies how computational modelling can contribute to the integration of observational and experimental radiation protection research.

## INTRODUCTION

Applying the two-stage clonal expansion model to epidemiology of lung cancer among uranium miners, it has been revealed that radon acts as a promoting agent facilitating the clonal expansion of already mutated cells. Clonal expansion rate increases non-linearly by radon concentration showing a plateau above ~150 WLM/year (1). The underlying mechanisms remain unclear. As inhaled radon progeny deposit in the lung airways highly heterogeneously (particle concentration can be few hundred times higher in the carina regions of the bronchial airways, than in their other parts (2)), and alpha-particles effectively kill cells (3), there is a strong need for increased cell production rate in the deposition hot spots in order to maintain the cell number in the tissue. There are two ways for that, both potentially affecting the clonal expansion rate: the increase of either the number or the division rate of progenitor cells (or both). Besides theoretical considerations (4), there are experimental (5) and histological data (6) suggesting that the number of progenitor cells increases upon chronic irritation. Therefore, earlier we proposed that increase in epithelial progenitor or basal cell number (hereinafter referred to as hyperplasia) occurs in the large bronchi upon chronic radon exposure (7,8). The objective of the present study is to estimate the potential consequences of basal cell hyperplasia on the relationship between clonal expansion rate and radon concentration applying numerical modelling techniques.

## METHODS

The main assumption of the study is that the clonal expansion rate estimated by mathematical analysis of epidemiological data (1) is proportional to cell division rate for all exposure rates. Therefore the dependence of cell division rate, $\alpha$ on exposure rate, $D$ can be described by the same function as the dependence of clonal expansion rate, $\gamma$ on exposure rate (Equation 1 and 2):

$$\gamma = \left(1 + r_1(1 - e^{-r_2 D / r_1})\right) \cdot \gamma_b = f(D) \cdot \gamma_b, \quad (1)$$

where $\gamma_b$ is the "normal" clonal expansion rate if there is no exposure, while $r_1 = 0.89$ and $r_2 = 0.019$ y/WLM are fitted parameters in (1). The same relationship means that

$$\alpha = f(D) \cdot \alpha_b, \quad (2)$$

where $\alpha_b$ is the "normal" division rate of basal cells in an unexposed tissue.

We intend to test whether we can find a link between data obtained from epidemiology and radiation biology.

*Corresponding author: balazs.madas@energia.mta.hu

Using Equation 2, cell division rate is estimated based on epidemiological data. However, cell division rate can be estimated also by modelling the biological response at the tissue level applying numerical epithelium and microdosimetry models similarly to our earlier works (7,8). We test whether we can find a reliable measure of hyperplasia for any given exposure rate which results in the same cell division rate as determined by Equation 2.

In the model of the biological response, cell division rate was obtained by supposing equilibrium between cell death and cell division. The spontaneous death rate of non-basal cells was obtained by the number, 17,100 per mm$^2$ (9) and division rate, varying between 1/7 and 1/100 day$^{-1}$ (10) of basal cells in an unexposed tissue. Survival probability, $p_{sv}$ upon alpha-exposure decreases exponentially by cell nucleus hits, $n$:

$$p_{sv} = e^{-\beta \cdot n}, \qquad (3)$$

where $\beta$ is 0.285 (3). The same function was applied for survival of basal and non-basal cells, although only basal cells were considered as progenitors in this study.

In order to calculate cell nucleus hits, numerical epithelium models were prepared with different number of basal cells corresponding to different measures of hyperplasia (Table 1), and were applied in an own-developed Monte-Carlo code presented in (7). It was supposed that cell volumes and the area of the epithelium did not change, and therefore the additional basal cells result in an increase in the epithelium thickness. Besides basal cells, five other types were considered: goblet, other secretory, ciliated, preciliated, and intermediate cells. The volumes, frequencies and depth distributions of cell nuclei were based on experimental data (9,11). If experimental data are not available in (9) and (11), the same assumptions are used as in (7). Cell nuclei were represented as spheres and were located in a rectangular cuboid with 400 μm × 400 μm basic area representing a small part of the bronchial epithelium. 50 different numerical epithelium models were generated for each measure of hyperplasia (Table 1) in order to estimate empirical standard deviations.

**Table 1. Properties of numerical epithelium models characterizing the tissue with different number of basal cells, i.e. different measures of hyperplasia.**

| Relative increase in basal cell number [%] | Total basal cell number per unit surface [mm$^{-2}$] | Thickness of the epithelium [μm] |
|---|---|---|
| 0 % | 17,100 | 57.80 |
| 50 % | 25,650 | 63.12 |
| 100 % | 34,200 | 68.45 |
| 150 % | 42,750 | 73.77 |
| 200 % | 51,300 | 79.10 |
| 250 % | 59,850 | 84.42 |
| 300 % | 68,400 | 89.75 |
| 350 % | 76,950 | 95.07 |
| 400 % | 85,500 | 100.40 |
| 450 % | 94,050 | 105.72 |

Only alpha-particles emitted by radon progeny were taken into account in the Monte-Carlo code. 10.4% of alpha-particles were produced by $^{218}$Po (6.00 MeV) and 89.6% by $^{214}$Po (7.69 MeV). The same isotope ratio was applied inside and outside of the deposition hot spots, and alpha-particles crossing the airway lumen were neglected. The bronchial epithelium is covered by a periciliary liquid layer and a mucus layer, which hereinafter collectively named as mucus layer with a total normal thickness of about 11 μm. The distribution of alpha-decays within the mucus layer was supposed to decrease exponentially by depth with a half-value thickness of 6 μm. The tracks of alpha-particles were considered as straight lines with isotropic direction distribution. The ranges of alpha-particles were determined by using the Stopping and Range of Ions in Matter (SRIM) software (12).

The relationship between the number of alpha-decays per unit surface per day and exposure rate was obtained from earlier studies. In the most exposed hot spot of 0.14 mm$^2$, 0.047 WLM (working level month[1]) equivalent to 8 h of work in a mine environment characterised by an exposure rate of 1 WL (working level) results in approximately 0.125 μm$^{-2}$ alpha-decay per unit surface if mucociliary clearance is neglected (7). However, the average number of alpha-decays per unit surface in the large bronchi is only $1.98 \times 10^{-4}$ μm$^{-2}$ upon the same 0.047 WLM exposure (8 h of work in mine with an exposure rate of 1WL) (13). This latter number was applied in simulations for the epithelium outside of the carina regions. 50 simulations were performed for each exposure rate, applying a different piece of the epithelium models.

---

[1]Working level month (WLM) is the historical unit of exposure to radon progeny applied to uranium mining environment. One WLM of exposure corresponds to breathing an atmosphere at a concentration of one working level (WL) for a working month of 170 h and it is equivalent to $3.54 \times 10^{-3}$ Jh m$^{-3}$. A concentration of 1 WL is any combination of short-lived radon progeny in one liter of air what will emit $1.3 \times 10^5$ MeV of alpha-particle energy.

We applied these models for different surface activities characteristic of different exposure rates in order to find the relationship between the measure of hyperplasia and exposure rate that results in the division rate-exposure rate function described by Equation 2. For example, Figure 1 shows functions between division rate and measure of hyperplasia for different exposure rates (curves with symbols). The division rates calculated from Equation 2 are plotted as horizontal lines without symbols. For each exposure rate, the intersection of the corresponding curves (with and without symbols) provides one data point in Figure 2. For example, the cell division rate for exposure rate of 30 WLM/y is 0.23 day$^{-1}$ (horizontal solid line), and the division rate obtained by our models is plotted as a decreasing solid curve with square symbols. Their intersection is found where the relative increase in basal cell number is 340%, which is plotted in Figure 2 at 30 WLM/y and 0.23 day$^{-1}$.

The relationship between division rate and exposure rate strongly depends on different parameters like mucus thickness (normal 11 μm and doubled 22 μm), normal cell division rate (1/7 day$^{-1}$ and 1/100 day$^{-1}$) (10), and the location in the large bronchi (inside or outside of the deposition hot spots). Therefore we performed simulations with different combinations of the parameters.

is 11 μm. It was found that a very strong increase in basal cell number is required to obtain the division rate computed from Equation (2) even if the exposure rate is just 30 WLM/y (Figure 2). It is not expected either that the basal cell number will be more than five times higher upon an exposure rate of around 110 WLM/y even if it resulted in a local dose of about 8.5 Gy per day in an epithelium with normal thickness (7).

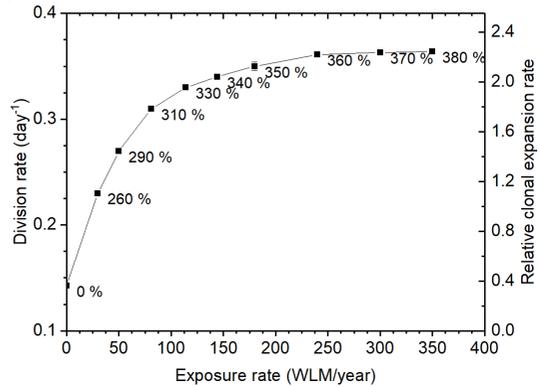

Figure 2. Division rate as the function of exposure rate in deposition hot spots if the normal division rate is 1/7 day$^{-1}$ and the mucus layer thickness is 11 μm. The relative increase in basal cell number required to obtain the division rate from Equation 2 can be seen next to the calculated points.

In response to different drugs and irritants, not only the number of progenitor cells but also the mucus thickness can increase (14). Therefore simulations for mucus layer thickness of 22 μm were also performed. In this case, the measure of hyperplasia required to meet the division rate obtained from Equation (2) is lower than in case of normal mucus thickness, as it can be seen in Figure 3 compared to Figure 2. However, a fourfold increase in basal cell number is still improbably high.

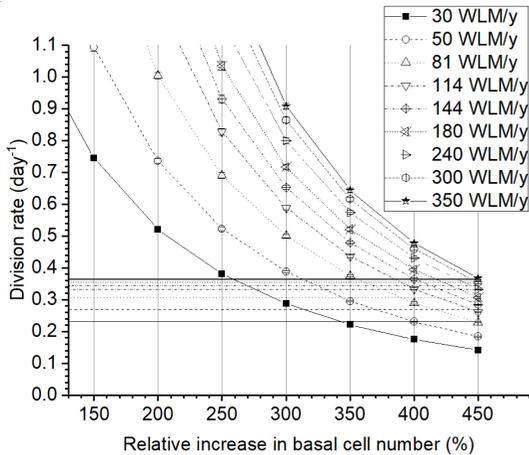

Figure 1. Division rate as the function of basal cell number in the deposition hot spots if the normal division rate is 1/7 day$^{-1}$ and the mucus layer thickness is 11 μm. The different curves with symbols refer to different exposure rates and the different horizontal lines refer to division rate obtained from Equation 2.

## RESULTS

First, the deposition hot spots in the bronchial airways were studied. Figure 1 shows the division rate as the function of the measure of hyperplasia if the normal division rate is 1/7 day$^{-1}$ and the mucus layer thickness

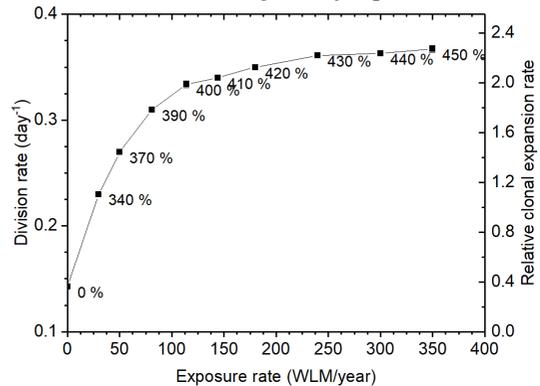

Figure 3. Division rate as the function of exposure rate in a deposition hot spot if the normal division rate is 1/7 day$^{-1}$ and the mucus layer thickness is doubled (22 μm). The relative increase in basal cell number required to obtain the division rate from Equation 2 can be seen next to the calculated points.

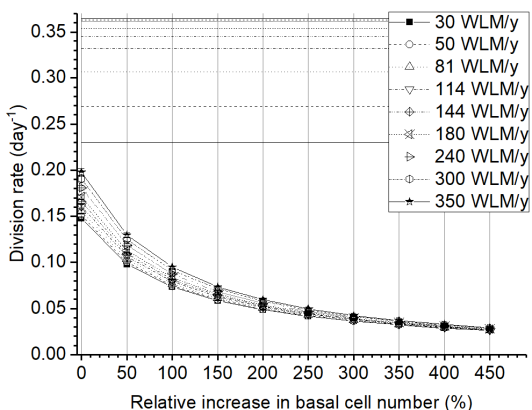

Figure 4. Division rate as the function of basal cell number outside the deposition hot spots if the normal division rate is 1/7 day$^{-1}$ and the mucus layer thickness is 11 μm. The different curves with symbols refer to different exposure rates and the different horizontal lines refer to division rate obtained from Equation 2.

Based on the results shown in Figure 2 and 3, the measure of hyperplasia required to reduce cell division rate to the level determined by Equation 2 is very high in the deposition hot spots. Hereinafter, therefore, we focus on the less exposed parts of the bronchial epithelium. Figure 4 shows cell division rate outside the deposition hot spots if the normal division rate is 1/7 day$^{-1}$ and the mucus layer thickness is 11 μm. In this case, even the highest simulated exposure rate results in a lower cell division rate than computed from Equation 2 which is further decreased by the measure of hyperplasia.

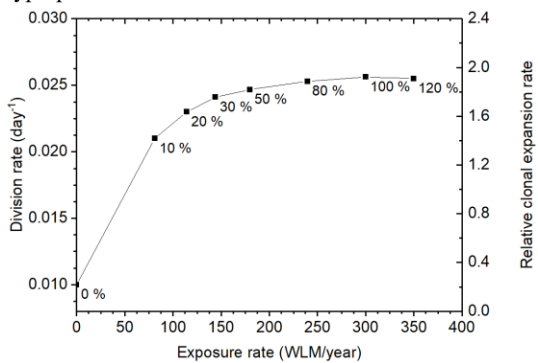

Figure 5. Division rate as the function of expansion rate outside the deposition hot spots if the normal division rate is 1/100 day$^{-1}$ and the mucus layer thickness is 11 μm. The relative increase in basal cell number required to obtain the division rate from Equation 2 can be seen next to the calculated points.

However, if the mucus layer thickness remains 11 μm, but the normal division rate is 1/100 day$^{-1}$, then there are intersections of cell division rates determined in different ways again. Therefore, there is a relationship between exposure rate and the measure of hyperplasia that results in the same exposure rate dependence of cell division rate as obtained from Equation 2. In addition, as Figure 5 shows, the required measure of hyperplasia is much lower than previously: 2.2-fold increase in basal cell number is enough for the tissue to cope with a 350 WLM/y exposure rate outside the deposition hot spots even with normal mucus thickness (11 μm).

As the required measure of hyperplasia is reasonable low in this latter case, we plotted the relationship between the measure of hyperplasia and exposure rate in Figure 6. It can be seen that basal cell number increases sublinearly by exposure rate below 200 WLM/y.

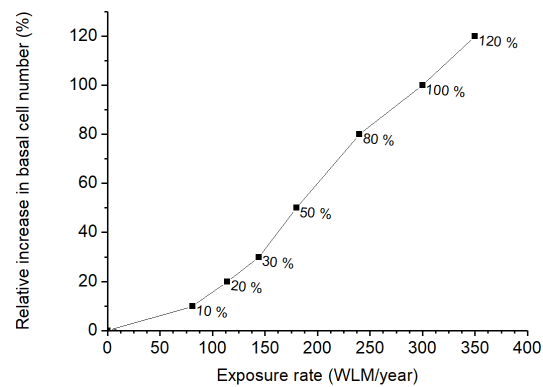

Figure 6. The measure of hyperplasia required as the function of exposure rate outside the deposition hot spots if the normal division rate is 1/100 day$^{-1}$ and the mucus layer thickness is 11 μm.

CONCLUSIONS

The objective of the present study was to test whether the induction of hyperplasia, which measure is dependent on exposure rate provides a quantitative explanation for the levelling in clonal expansion rate as observed in mathematical analysis of epidemiological data (1). As the applied numerical models have several parameters measured experimentally in a wide range, we searched the parameter space for an appropriate combination. We found a parameter setting where there is a reasonable dependence of the measure of hyperplasia on exposure rate to result in the same relationship between cell division rate and exposure rate as it is between clonal expansion rate and exposure rate. However, one of the key parameters, the cell division rate in an unexposed tissue is quite uncertain varying between 1/7 and 1/100 day$^{-1}$ because of the difficulty of measurement as it is discussed in Adamson's work (10).

Therefore further studies are required to find a conclusive answer on the question whether the induction of hyperplasia is responsible for the levelling in clonal expansion rate, and for the corresponding inverse exposure rate effect (15). Histological studies focusing on the distribution of tissue thickness and basal cell number along the bronchial airways of former uranium miners can provide crucial insight on the role of potential changes in tissue architecture, like basal cell hyperplasia in radon carcinogenesis.

ACKNOWLEDGMENTS

This work was supported by the European Union and the State of Hungary, co-financed by the European Social Fund in the framework of TÁMOP 4.2.4. A/2-11-1-2012-0001 "National Excellence Program" (A2-EPFK-13-0160), and by the National Research, Development and Innovation Office (VKSZ_14-1-2015-0021).